\documentclass[10pt]{article}

\usepackage{amssymb}
\usepackage{color}
\usepackage{epsfig,amssymb,amsfonts,amsmath,graphicx,dsfont,cite,xfrac}
\usepackage{authblk}
\usepackage{subcaption}
\definecolor{mygray}{gray}{0.5}

\usepackage{cite}
\usepackage[colorlinks=true,linkcolor=blue,citecolor=red]{hyperref}
\title{Coherent control of two Jaynes-Cummings cavities}

\author[1]{L. O. Casta\~nos-Cervantes}
\author[2]{Lorenzo M. Procopio}
\author[3,*]{Marco Enr\'iquez}

\affil[$^{1}$]{\footnotesize Tecnologico de Monterrey, School of Engineering and Sciences, Ciudad de Mexico 14380, Mexico}
\affil[$^{2}$]{\footnotesize Weizmann Institute of Science, Rehovot 7610001, Israel}
\affil[$^{3}$]{\footnotesize Tecnologico de Monterrey, School of Engineering and Sciences, Santa Fe 01389, Mexico}

\affil[*]{menriquezf@tec.mx}

\begin{document}

\maketitle

\begin{abstract}
In this work, we uncover new features on the study of a two-level atom interacting with one of two cavities in a coherent superposition. The James-Cummings model is used to describe the atom-field interaction and to study the effects of quantum indefiniteness on such an interaction. We show that coherent control of the two cavities in an undefined manner allows novel possibilities to manipulate the atomic dynamics on demand which are not achievable in the conventional way. In addition, it is shown that the coherent control of the atom creates highly entangled states of the cavity fields taking a Bell-like or Schr\"odinger-cat-like state form. Our results are a step forward to understand and harness quantum systems in a coherent control, and open a new research avenue in the study of atom-field interaction exploiting quantum indefiniteness.
\end{abstract}

%\flushbottom
%\maketitle
% * <john.hammersley@gmail.com> 2015-02-09T12:07:31.197Z:
%
%  Click the title above to edit the author information and abstract
%
%\thispagestyle{empty}

%\noindent Please note: Abbreviations should be introduced at the first mention in the main text – no abbreviations lists. Suggested structure of main text (not enforced) is provided below.

\section{Introduction}

The Jaynes-Cummings (JC) model is one of the fundamental models used to describe the interaction between light and matter \cite{jaynes1963comparison}. It describes the interaction of a two-level atom (or {\it qubit}) with a single-mode quantum electromagnetic field when both the detuning between the atom's transition frequency and the field's frequency and the atom-field coupling are much smaller than the field's frequency \cite{shore1993jaynes,larson2022jaynes}. For example, it has been applied successfully in cavity quantum electrodynamics (QED) \cite{walther2006cavity,Haroche}. However, the interaction of one atom with one of two cavities in an undefined manner is largely unexplored. 

In this context, a novel technique has been proposed to coherently control the order of quantum operations in the frame of quantum computing \cite{Chiribella2013}. In relation with quantum communications, this method creates an indefiniteness in the order of application of two \cite{ebler2018enhanced}  or more successive quantum channels \cite{procopio2019communication}. Furthermore, new quantum advantages in quantum computation \cite{taddei2021computational,renner2022computational}, quantum communication complexity \cite{feix2015quantum,guerin2016exponential}, quantum metrology \cite{zhao2020quantum,procopio2022parameter}, and quantum thermodynamics \cite{felce2020quantum,simonov2022work} have been reported. Several experiments have been performed to show those advantages \cite{wei2019experimental,taddei2021computational,nie2020experimental}. A simpler type of indefiniteness can be created by just placing the quantum system of interest in a coherent superposition of two alternative locations \cite{oi2003interference,ban2022temporal,ban2020decoherence}. In this technique one has control over the choice on which path the quantum system will go through \cite{aharonov1990superpositions} achieving new quantum advantages in quantum communications \cite{abbott2020communication}, quantum coherence  \cite{ban2020decoherence}, and quantum metrology \cite{chiribella2022heisenberg}.

Motivated by this research, we propose to coherently control a two-level atom interacting with two quantum cavity fields in a superposition of two different spatial locations. Recently, it has been proposed to use quantum indefiniteness in the order of application of two cavities following the Jaynes-Cummings model \cite{fellous2023comparing}. However,  they study quantitatively the energetic differences between  different strategies rather than to study the effects of indefiniteness per se on the atom-field interaction. We show that new interesting effects are unveiled applying indefiniteness to the interaction of one atom with two cavity fields. 
For example, dealing with the atomic population, the path superposition gives rise to novel intriguing features in the atom-field interaction not present in the conventional case. 

To determine which cavity the atom will interact with we use a control qubit encoding the spatial path of the atom. If the control qubit is in state  $\vert 0\rangle$, the atom will interact with the electromagnetic field in cavity $C_0$, Likewise, if the control qubit is in state  $\vert 1\rangle$, the atom will interact with the electromagnetic field in cavity $C_1$.  By sending the control qubit in a superposition of its quantum states $\vert \theta,\varphi \rangle=\cos \theta \vert0\rangle+e^{i\varphi}\sin \theta \vert1\rangle$, we coherently superpose both cavity fields and maximum indefiniteness is achieved when $\theta=\pi/4$. 

In this work we report some contributions in the study of one qubit interacting with two cavity fields in a coherent superposition. To show the usefulness of the method, we focus on two aspects of the effects of indefiniteness on the atom-field interaction: the effects on the inversion of population of the atom and the effects on the cavity fields. In the first case, the atom enters the cavities and the dynamics of the system are determined while the atom traverses them. Here the cavity fields are supposed to be described with a definite number of photons. The second case of study is similar to Young's double slit experiment, since the atom goes through both cavities, interacts dispersively with the cavity fields which are initially described by coherent states, and then exits the cavities. We make two different types of measurements on the whole system. One consists in measuring the state of the control qubit, while the other consists in measuring the state of the atom. We found that our method creates highly entangled states of the cavity fields that can take a Bell-like or Schr\"{o}dinger cat form. Moreover, there can be a nonnegligible probability to find both cavity fields in Schr\"{o}dinger cat states.

%The article is organized as follows. In Section II we present our results for the first case of study. Section III reviews the JC model in the dispersive regime and Section IV  presents our second case of study. Finally, the conclusions are in Section V.

%%%%%%%----->Section

\section{The Jaynes-Cummings model}

The JC model describes a system composed of a qubit (a quantum two-level system) interacting with a harmonic oscillator and it is obtained from the Rabi model \cite{PRA,PRAa} by applying the rotating wave approximation (RWA). The JC Hamiltonian is
\begin{eqnarray}
\label{1}
H_{\mbox{\tiny JC}} = \frac{\hbar \omega_{a}}{2} \sigma_{z} + \hbar \omega a^{\dagger}a + \hbar g (\sigma_{-}a^{\dagger} + \sigma_{+}a) \ ,
\end{eqnarray}
where $\omega_{a}>0$ is the angular transition frequency of the qubit, $\omega >0$ is the angular frequency of the harmonic oscillator, and $g$ is a real number with units $1/s$ that describes the strength of the qubit-oscillator coupling. For simplicity, here and in the following we omit the energy of the ground state of the oscillator.

An orthonormal basis for the state space of the qubit is $\left\{ \ \vert e \rangle , \ \vert g \rangle \ \right\}$ where $\vert e \rangle$ and $\vert g \rangle$ denote the excited and ground states of the qubit, respectively. Also, the qubit raising and lowering operators are respectively given by
\begin{eqnarray}
\sigma_{+} &=& \vert e \rangle\langle g \vert , \quad \sigma_{-} = \vert g \rangle\langle e \vert ,
\end{eqnarray}
and $\sigma_{x}$, $\sigma_{y}$, and $\sigma_{z}$ denote the Pauli operators defined by
\begin{eqnarray}
\sigma_{x} &=& \sigma_{-} + \sigma_{+} \ , \quad \sigma_{y} = i (\sigma_{-} - \sigma_{+}) \ , \quad
\sigma_{z} = \vert e \rangle\langle e \vert - \vert g \rangle\langle g \vert \ . 
\end{eqnarray}
In addition, $a^{\dagger}$ and $a$ are the creation and annihilation operators of the oscillator. The harmonic oscillator usually represents a single-mode of the electromagnetic field, while the qubit is a two-level real or artificial atom. 
Since the JC model is obtained by applying the RWA, it requires a small qubit-oscillator coupling and a small qubit-oscillator detuning with respect to the qubit and oscillator frequencies, that is, it requires
\begin{eqnarray}
\label{RWA}
\vert g \vert , \ \vert \Delta \vert &\ll& \omega_{a} + \omega \ ,
\end{eqnarray}
where the detuning $\Delta$ is defined as
\begin{eqnarray}
\label{detuning}
\Delta &=& \omega_{a} - \omega \ .
\end{eqnarray}
The excitation number operator
\begin{eqnarray}
\label{excitationOP}
N &=& a^{\dagger}a + \frac{1}{2}\sigma_{z} \ ,
\end{eqnarray}
is a constant of the motion for $H_{\mbox{\tiny JC}}$ and one can write
\begin{eqnarray}
H_{\mbox{\tiny JC}} &=& \hbar \omega N + \hbar V \ ,
\end{eqnarray}
where $N$ conmutes with $V$ and
\begin{eqnarray}
\label{V}
V &=& \frac{\Delta}{2} \sigma_{z} + g(\sigma_{-}a^{\dagger} + \sigma_{+}a) \ . 
\end{eqnarray}
Then, the evolution operator associated with $H_{\mbox{\tiny JC}}$ can be expressed as
\begin{eqnarray}
\label{opEv}
e^{-\frac{i}{\hbar}H_{\mbox{\tiny JC}}(t-T_{0})} &=& e^{-i\omega N (t-T_{0})} e^{-iV(t-T_{0})} \ ,
\end{eqnarray}
In the following we use the JC model to describe the atom-field interaction of the physical system under study.
%%%%%%----->Section

\section{The system under study}

We consider a system composed of a control qubit, a two-level atom (we use the same notation as the previous section), and two single-mode quantum electromagnetic fields with angular frequencies $\omega_{0}$, $\omega_{1} >0$. The field with frequency $\omega_{k}$ is contained in cavity $k$. At time $t = 0$ the state of the system is prepared and the atom is shot towards the cavities. It moves with constant velocity, enters both cavities at a time $t=T_0$, and interacts with them during a time-interval.

\begin{figure}[h]
\centering
\begin{tabular}{cc}
\includegraphics[scale=0.37]{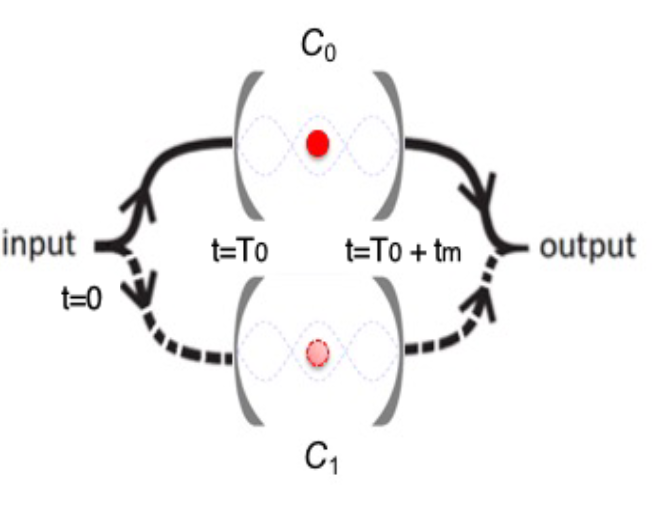} & \includegraphics[scale=0.25]{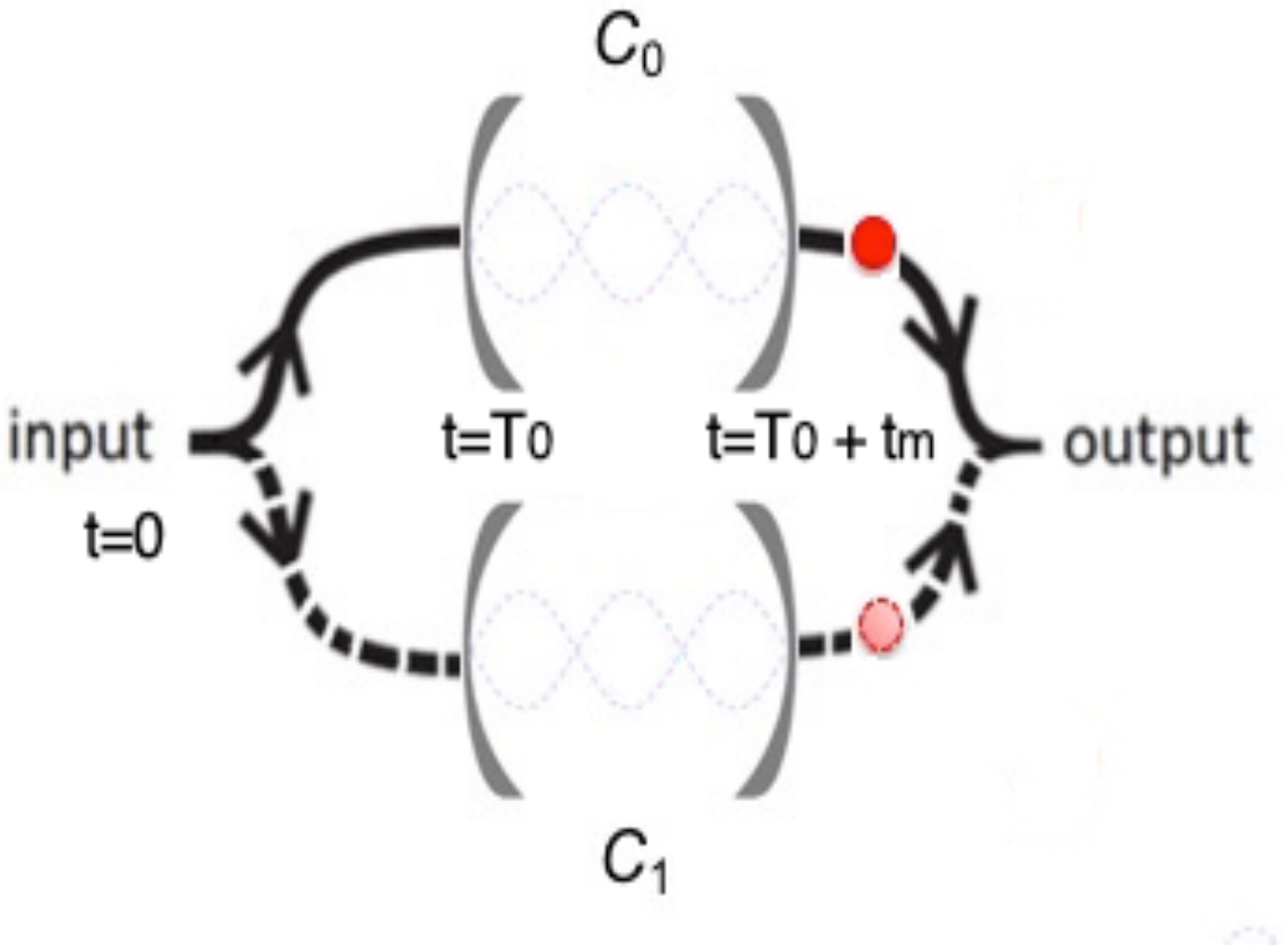}\\
(a) & (b)
\end{tabular}
\caption{\label{Figure0} The figure depicts the system under consideration. At time $t=0$ the state of the system is prepared and, in the first stage, the atom moves with constant velocity reaching the cavities at the time $t=T_{0}$. (a) In the first situation we consider that the atom interacts with both cavity fields during a certain time interval. (b) In the second scheme an additional stage is considered: the atom exits the cavities.}
\end{figure}

In this article we consider two scenarios. In the first one, a projective measurement is performed and some physical quantities are measured while the atom is inside the cavities. In the second one, the atom traverses the cavities, exits them, and then we study how entanglement between the the two cavity fields can be created. This last physical situation is similar to Young's double slit experiment. These cases are schematically illustrated in Figure \ref{Figure0}.

The control qubit determines the path the atom goes through. In the following, the sub-index $c$ is used to identify quantities associated with the control qubit. An orthonormal basis for the state space of the control qubit is $\left\{ \ \vert 0 \rangle_{c} , \ \vert 1 \rangle_{c} \ \right\}$. If the control qubit is in the state $\vert k\rangle_{c}$, then the atom passes only through cavity $k$ ($k=0,1$). Thus, the superposition 
\begin{eqnarray}
\label{Control}
\vert \theta,\varphi \rangle_{c} &=& \cos \theta \vert0\rangle_{c} + e^{i\varphi}\sin \theta \vert1\rangle_{c},
\end{eqnarray}
implies that the atom passes through both cavities. The probability of the atom going through the cavity 0 or 1 is $\mbox{cos}^{2}(\theta)$ or  $\mbox{sin}^{2}(\theta)$, respectively.
In addition, $0\le  \theta <\pi/2$ and $ 0\leq \phi < 2\pi$. Setting the parameters $\theta=\pi/4$ and $\varphi=0$ we define the state
\begin{eqnarray}
\label{Control}
\vert + \rangle_{c} &=& \frac1{\sqrt 2}\left(\vert0\rangle_{c} + \vert1\rangle_{c}\right),
\end{eqnarray}
for which the maximum indefiniteness is achieved. Alternatively, the state
\begin{eqnarray}
\label{Controlm}
\vert - \rangle_{c} &=& \frac1{\sqrt 2}\left(\vert0\rangle_{c} - \vert1\rangle_{c}\right),
\end{eqnarray}
corresponds to the parameters $\theta=\pi/4$ and $\varphi=\pi$.
%{\color{red} 
Outside the cavities the system Hamiltonian is given by the atomic and field free energies as
\begin{equation}\label{hfree}
 H_{\mbox{\tiny free}} = \frac{\hbar\omega_{a}}{2}\sigma_{z} + \hbar\omega_{0} a_{0}^{\dagger}a_{0} + \hbar\omega_{1} a_{1}^{\dagger}a_{1}
\end{equation}
where cavity field $k$ has creation and annihilation operators given by $a_{k}^{\dagger}$ and $a_{k}$, respectively. When the atom interacts with a superposition of both cavities the Hamiltonian is
\begin{eqnarray}\label{hinteraction}
 H_{I} &=& \vert 0 \rangle_{cc}\langle 0 \vert\otimes \Big( H_{\mbox{\tiny JC}}^{(0)} + \hbar\omega_{1} a_{1}^{\dagger}a_{1} \Big) + \vert 1 \rangle_{cc}\langle 1 \vert\otimes \Big( H_{\mbox{\tiny JC}}^{(1)} + \hbar\omega_{0} a_{0}^{\dagger}a_{0} \Big),
\end{eqnarray}
where $H_{\mbox{\tiny JC}}^{(k)}$ is the usual JC Hamiltonian for the atom and cavity field $k$, that is,
\begin{equation}
 H_{\mbox{\tiny JC}}^{(k)} = \frac{\hbar \omega_{a}}{2} \sigma_{z} + \hbar \omega_{k} a_{k}^{\dagger}a_{k} + \hbar g_{k} (\sigma_{-}a_{k}^{\dagger} + \sigma_{+}a_{k}), \quad k=0,1.
\end{equation}
Here $g_{k}$ is a real number with units $1/s$ that denotes the coupling of the atom with cavity field $k$.
Using (\ref{excitationOP})-(\ref{V}) the evolution operator of the system during the time interval $[T_0, T_0+t]$ is
\begin{eqnarray}
\label{op1}
e^{-\frac{i}{\hbar}H_{I}(t-T_{0})} &=& \vert 0 \rangle_{cc}\langle 0 \vert \otimes e^{-\frac{i}{\hbar}H_{\mbox{\tiny JC}}^{(0)}(t-T_{0})}e^{-i\omega_{1}a_{1}^{\dagger}a_{1}(t-T_{0})}  + \vert 1 \rangle_{cc}\langle 1 \vert \otimes e^{-\frac{i}{\hbar}H_{\mbox{\tiny JC}}^{(1)}(t-T_{0})}e^{-i\omega_{0}a_{0}^{\dagger}a_{0}(t-T_{0})}  .
\end{eqnarray}
Using (\ref{opEv}) one has
\begin{eqnarray}
\label{OpEvF}
e^{-\frac{i}{\hbar}H_{\mbox{\tiny JC}}^{(k)}(t-T_{0})} 
&=& e^{-i\omega_{k} N_{k} (t-T_{0})} e^{-iV_{k}(t-T_{0})} \ .
\end{eqnarray}
Here, the excitation number operator $N_{k}$ for cavity field $k$, the operator $V_{k}$, and the detuning $\Delta_{k}$ with the frequency of cavity field $k$ are given by
\begin{eqnarray}
\label{VF}
N_{k} &=& a_{k}^{\dagger}a_{k} + \frac{1}{2}\sigma_{z} \ , \cr
V_{k} &=& \frac{\Delta_{k}}{2} \sigma_{z} + g_{k}(\sigma_{-}a_{k}^{\dagger} + \sigma_{+}a_{k}) \ , \cr
\Delta_{k} &=& \omega_{a} - \omega_{k} . 
\end{eqnarray}
Observe that $N_{k}$ and $V_{k}$ commute. Then, the evolution operator can be expressed as
\begin{eqnarray}
\label{opEVfinal}
e^{-\frac{i}{\hbar}H_{I}(t-T_{0})} 
= U_{IF}(t,T_0)\Bigg[\vert 0 \rangle_{cc} \langle 0 \vert\otimes  W_0(t,T_0) +\vert 1 \rangle_{cc} \langle 1 \vert\otimes W_1(t,T_0) \Bigg] , \quad
\end{eqnarray}
where $ W_k(t,T_0)=e^{-iV_{k}(t-T_{0})}$,
and we have introduced the unitary operator
\begin{equation}
\label{IPF}
U_{IF}(t,T_0) = \vert 0 \rangle_{cc}\langle 0 \vert\otimes \Lambda_0(t,T_0)  + \vert 1 \rangle_{cc} \langle 1 \vert\otimes \Lambda_1(t,T_0) ,
\end{equation}
with
\begin{equation}
 \Lambda_k(t,T_0)=e^{-i\omega_{k}N_{k}(t-T_{0})} e^{-i\omega_{k\oplus1}a_{k\oplus1}^{\dagger}a_{k\oplus1}(t-T_{0})},
\end{equation}
where $\oplus$ stands for the sum$\mod 2$. Finally, according to \cite{Kli09} the operator $W_k(t,T_{0})$ can be expressed as
\begin{equation}
\label{Klimov}
%\begin{array}{ll}
W_k(t,T_0)=\displaystyle\cos \Omega_k \Big[ (N_k+1/2)(t-T_0) \Big] -i \frac{\sin \Big[\Omega_k(N_k+1/2)(t-T_0) \Big]}{\Omega_k(N_k+1/2)} V_k,
%\end{array}
\end{equation}
here we have introduced the quantity
\begin{eqnarray}
\Omega_{k}(x) &=& \sqrt{g_{k}^{2}x+\frac{\Delta_{k}^{2}}{4}} \ , 
\end{eqnarray}
where $x$ can be a real number or an operator.

%%%%%%----->Section

\section{Rabi oscillations}\label{Rabi}
In this section we focus on the effects of the indefiniteness of path on the state of the atom as it transits through the cavities. In order to describe the dynamics we analyze both the atomic population inversion and the photon number in each cavity. The initial state of the system is
\begin{eqnarray}
\label{psi0F}
\vert \psi (0) \rangle &=&  \vert \theta, \phi \rangle_{c}\otimes \vert e \rangle \otimes \vert n_0 \rangle_{0} \otimes \vert n_1 \rangle_{1} \ ,
\end{eqnarray}
that is, the control qubit is in the state (\ref{Control}), the atom is in the excited state $\vert e \rangle$, and cavity field $k$ is in the Fock state $\vert n_{k} \rangle_{k}$ ($n_{k}$ a no-negative integer). Hereafter the tensor product notation and the subsystem's indices will be omitted to simplify the notation.

We assume that the atom starts to interact with both cavity fields from $T_0=0$ onwards. Note that, before the atom enters the cavities, the corresponding time-evolution operator only adds a physically irrelevant global phase to (\ref{psi0F}) and, thus, can be omitted. The state of the system at some time $t_{m} >0$ is given by
\begin{eqnarray}
 \vert \psi(t_m)\rangle = e^{-\frac i\hbar H_I t_m} \vert \psi (0) \rangle \nonumber=\cos\theta \vert 0\rangle \otimes \Lambda_0 W_0 \vert e,n_0,n_1\rangle +e^{i\varphi} \sin\theta \vert 1\rangle \otimes \Lambda_1 W_1 \vert e,n_0,n_1 \rangle,
\end{eqnarray}
where the notation $A_j = A_j(t_m,0)$ is used for operators $\Lambda_j$ and $W_j$. At time $t=t_m$ a projective measurement on the control subsystem is performed, which is described by the projector $\vert \theta,\varphi\rangle\langle \theta,\varphi\vert$. 
Immediately after such measurement, the state of the system becomes
\begin{equation}
 \vert \psi'(t_m)\rangle=\frac{\vert \theta,\varphi \rangle}{{\cal N}_0} \otimes(\cos^2 \theta\Lambda_0W_0+\sin^2\theta \Lambda_1W_1)\vert e,n_0,n_1\rangle,
\end{equation}
where the normalization constant ${\cal N}_0$ is defined by 
\begin{equation}\label{n0}
\begin{array}{ll}
{\cal N}_0^2= \cos^4\theta+\sin^4\theta +2 \sin^2\theta\cos^2\theta  {\rm Re}(\langle e,n_0,n_1\vert W_0^\dagger \Lambda_0^\dagger\Lambda_1W_1\vert e,n_0,n_1\rangle).%\cr 
\end{array}
\end{equation}
Then, the state of the system at times $t\ge t_m$ is
$\vert\psi(t)\rangle = e^{-\frac i\hbar H_I(t-t_m)}\vert \psi'(t_m)\rangle$. Accordingly,
\begin{equation}\label{psidt}
 \vert \psi(t)\rangle = \frac{\vert 0\rangle}{{\cal N}_0}\otimes(\cos^3\theta T_{00}+\cos\theta \sin^2\theta T_{01})\vert e,n_0,n_1\rangle+\frac{\vert 1\rangle}{{\cal N}_0} \otimes(e^{i\varphi} \sin\theta \cos^2\theta T_{10} + e^{i\varphi} \sin^3\theta T_{11})\vert e,n_0,n_1\rangle,
\end{equation}
where $T_{ij}=\Lambda'_iW_i'\Lambda_jW_j$ for $i,j=0,1$ and the prime in each operator stands for $A_j' = A_j(t, t_m)$. 

\subsection{The atomic population inversion}
\begin{figure}[htbp]
\centering
\begin{tabular}{cc}
 \includegraphics[scale=0.5]{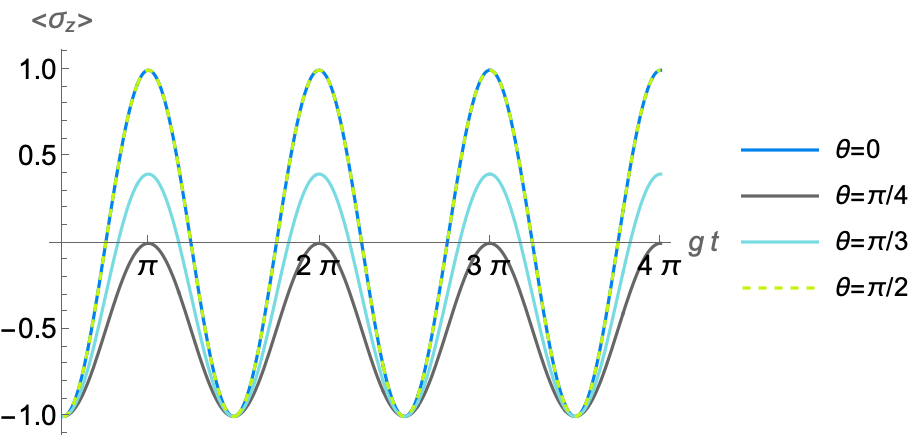} &  \includegraphics[scale=0.35]{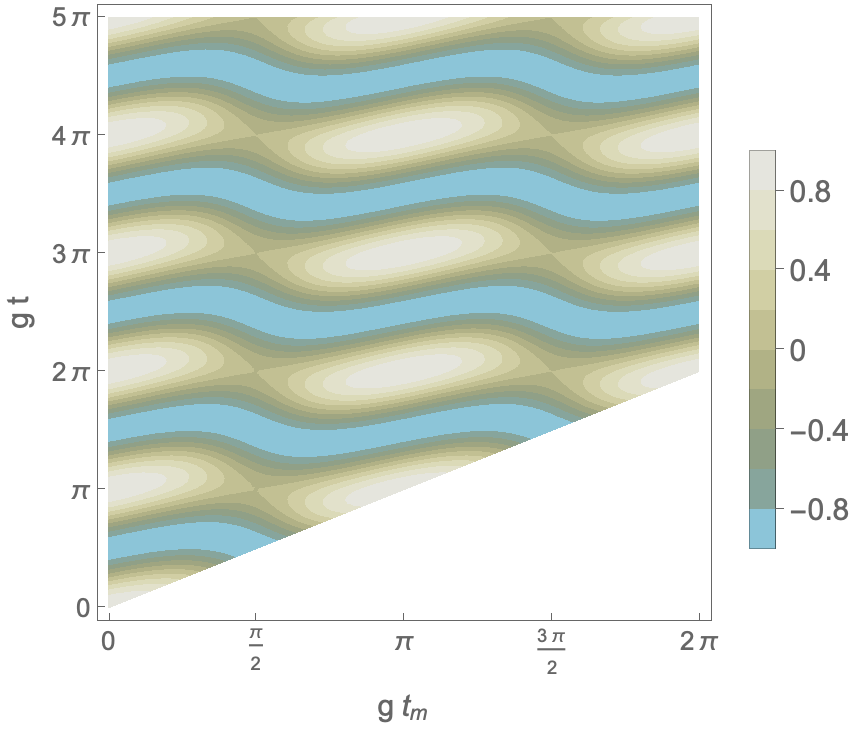}\\
 (a) & (b)
\end{tabular}
\caption{The atomic population inversion (\ref{sigmazgen}) time-evolution  when both cavity fields start out in the vacuum state, i.e., $n=0$ as function of some relevant parameters. (a) The effect of manipulating the control parameter $\theta$ with fixed value $t_m=\pi/(2 g)$. (b) The dependence of the population inversion on the measuring time $t_m$ for the fixed control parameter value $\theta=\pi/4$.}
\label{szn0}
\end{figure}

\begin{figure}[htbp]
\centering
\begin{tabular}{cc}
 \includegraphics[scale=0.55]{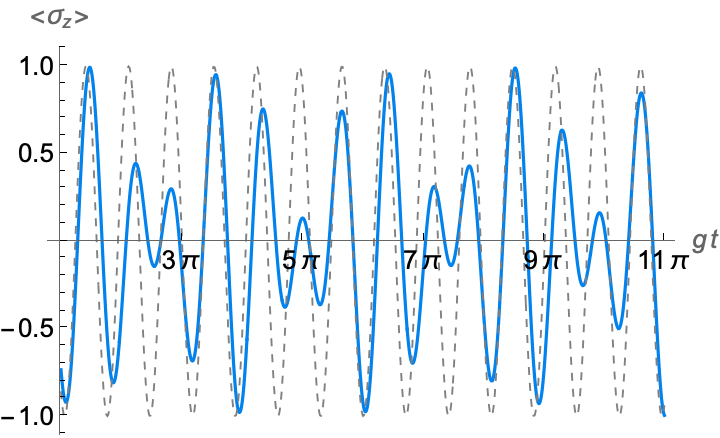} &  \includegraphics[scale=0.55]{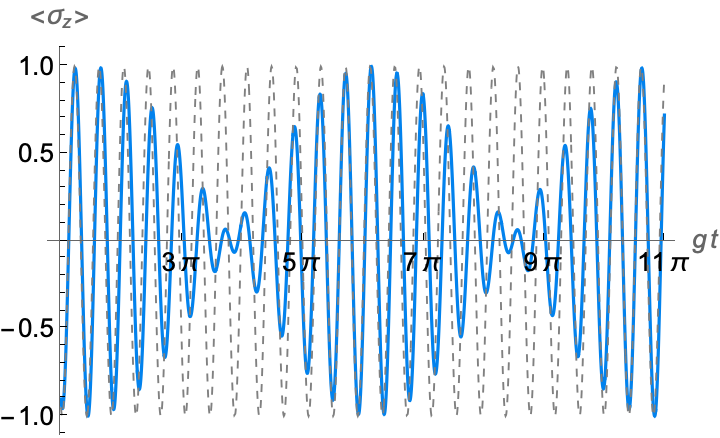}\\
 (a) & (b)
\end{tabular}
\caption{Time-evolution of the atomic population inversion (\ref{sigmazgen}) when the field in both cavities contains initially (a) one photon and (b) five photons (solid blue lines) with $t_m=\pi/g$. In addition, the dashed line corresponds to the single cavity atomic population inversion with the same photon number.}
\label{szn}
\end{figure}

We first analyze the dynamics of the atomic population inversion. A straightforward calculation shows that $\langle \sigma_z\rangle$ as function of time can be expressed as
\begin{eqnarray}\label{Invpop}
 \langle\sigma_z\rangle {\cal N}_0^2&=& \cos^6\theta \langle  T_{00}^\dagger \sigma_zT_{00} \rangle + \sin^6\theta \langle T_{11}^\dagger\sigma_zT_{11} \rangle%\cr
+2 \cos^4\theta \sin^2\theta {\rm Re} [ \langle T_{01}^\dagger\sigma_z T_{00}\rangle ]\cr
&& +2 \sin^4\theta \cos^2\theta {\rm Re} [ \langle T_{10}^\dagger \sigma_z T_{11}\rangle ] +\cos^2\theta \sin^4\theta \langle T_{01}^\dagger \sigma_zT_{01}\rangle +\sin^2\theta \cos^4\theta \langle T_{10}^\dagger \sigma_z T_{10}\rangle,
\end{eqnarray}
where each expectation value is computed on the state $\vert e,n_0,n_1\rangle$. Note that, in the case of no superposition, equation (\ref{Invpop}) reduces to the usual expression for a single cavity
\begin{equation*}
\langle \sigma_z \rangle = \langle  T_{jj}^\dagger \sigma_zT_{jj}\rangle=\frac{\Delta_j^2}{4\Omega_j^2(n_j)}+\left(1-\frac{\Delta_j^2}{4\Omega_j^2(n_j)}\right)\cos[2 \Omega_j(n_j) t],
\end{equation*}
where $j=0$ ($=1$) for $\theta = 0$ ($=\pi/2$).
Otherwise, there will exist interference as will be shown.
For the purposes of this study, it will be assumed that both cavities are in exact resonance, that is to say, $\Delta_0=\Delta_1=0$ and, hence, $\omega_0=\omega_1$. 
Then, the normalization constant (\ref{n0}) reduces to
\begin{equation}
{\cal N}_0^2= \cos^4\theta+\sin^4\theta+2 \sin^2\theta\cos^2\theta\cos(t_m g_0 \sqrt{n_0+1})\cos(t_m g_1 \sqrt{n_1+1}),
\end{equation}
and the expectation values in equation (\ref{Invpop}) are explicitly given as
\begin{equation}
{\rm Re} [ \langle T_{ji}^\dagger\sigma_z T_{jj}\rangle ] = \cos [(2t-t_m)g_j \sqrt{n_j+1}] \cos(t_m g_i\sqrt{n_i+1}),
\end{equation}
and
\begin{equation}
\langle T_{ji}^\dagger \sigma_zT_{ji}\rangle=
\cos[2(t-t_m)g_j \sqrt{n_j+1}]\cos^2(t_m g_i\sqrt{n_i+1})-\cos[2(t-t_m)g_j \sqrt{n_j}]\sin^2(t_m g_i\sqrt{n_i+1}),
\end{equation}
for $i\neq j$. In addition, we are interested in the identical cavities case to explore the effects of the superposition of paths, so we choose $n_0=n_1=n$ and $g_0=g_1=g$. In this case, the population inversion reads
\begin{equation}\label{sigmazgen}
\footnotesize
\begin{array}{ll}
\langle\sigma_z\rangle = \\[1ex] 
 \displaystyle \frac{2 (3+\cos 4\theta )\cos (2 gt\sqrt{n+1})+2 \sin^22\theta(\cos [2g(t-t_m) \sqrt{n+1}](1+\cos^2(gt_m\sqrt{n+1}))-\cos [2g(t-t_m) \sqrt{n}]\sin^2(gt_m\sqrt{n+1})}{7+\cos 4\theta+2\sin^2 2\theta \cos(2gt_m\sqrt{n+1})}.
\end{array}
\end{equation}
We first consider the case of zero photons in both cavities. No effect of the superposition is observed when the projective measurement time is $t_m={r \pi}/g$ with $r$ a non-negative integer as the expression (\ref{sigmazgen}) reduces to the single cavity population inversion regardless of the control parameter value. On the other hand, in Fig.~\ref{szn0}(a) we show the effect of the control parameter on the population inversion. Note that for the maximum indefinitess value, i.e., $\theta=\pi/4$ the probability of finding the atom in the ground state is always greater than the probability of finding in the excited one. Fig.~\ref{szn0}(a) also shows that the control parameter $\theta$ can be used to modify the population inversion amplitude on demand. Besides, Fig.~\ref{szn0}(b) depicts the time-evolution of the population inversion as function of the measurement time $t_m$ when the control state is given by $\vert+\rangle_c$. We note that the oscillation amplitude strongly depends on the $t_m$ value. For instance, the population inversion is always negative for $t_m=(2r+1)\pi/(2g)$, with $r$ a no-negative integer. 
On the other hand, in Fig.~\ref{szn} we depict the population inversion (\ref{sigmazgen}) for the non-vanishing photon number case. Two instances are considered and compared with the corresponding single cavity population inversion. We observe that the control parameter changes the uniform oscillatory behaviour noted in the conventional case. Besides, the plot shows that as $n$ increases an evolvent appears on the oscillations. 
\subsection{Photon number analysis}
\subsubsection*{Average photon number}
\begin{figure}[htbp]
\centering
\begin{tabular}{cc}
 \includegraphics[scale=0.4]{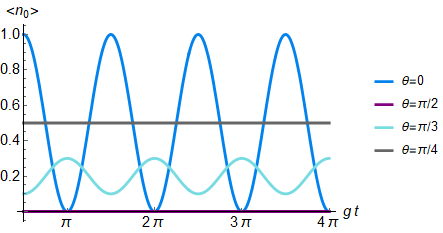} &  \includegraphics[scale=0.4]{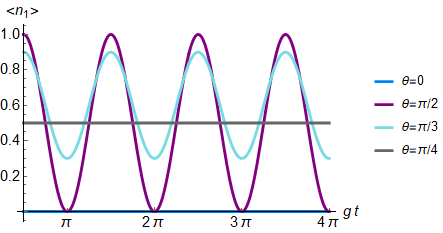}\\(a) &(b)\\
 \includegraphics[scale=0.42]{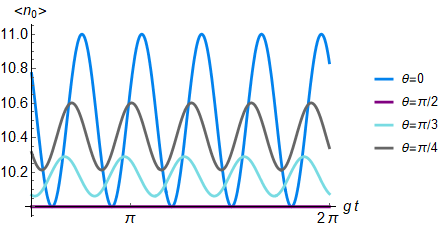} &  \includegraphics[scale=0.42]{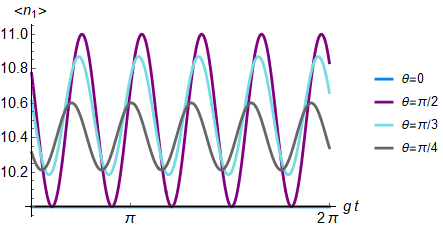}\\(c) &(d)\\
\end{tabular}
\caption{Average number of photons $\langle a_j^\dagger a_j \rangle$ for identical cavities, $n_0=n_1=n$ at different values of the control parameter $\theta$. We use equation \ref{n0nm} to plot the average number with the values $g_0=g_1=0.5$  and $t_m=\pi/2$ for two different cases of number of photons but equal to each cavity. Case 1 when $n_0=n_1=0$ photons for cavity 0, Figure (a) and for cavity 1, Figure (b). Case 2 when $n_0=n_1=10$ photons for cavity 0, Figure (c) and for cavity 1, Figure (d).}
\label{anp}
\end{figure}

\begin{figure}[htbp]
\centering
\begin{tabular}{cc}
\includegraphics[scale=0.6]{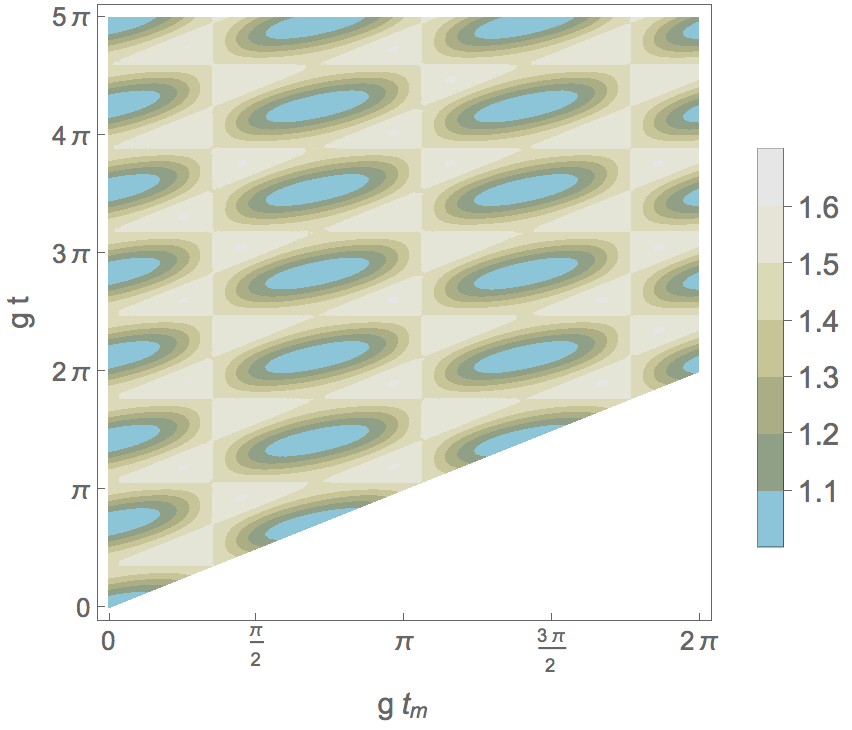}\\
\end{tabular}
\caption{The average number of photons $\langle n_j\rangle$  at maximum indefiniteness as function of the measurement $t_m$ and $t$ for $n_0=n_1=1$ and $g_0=g_1=0.2$.}
\label{noma}
\end{figure}

We also analyze the effects of indefiniteness on the number of photons in each cavity. First we analyze the average photon number  $\langle a_k^\dagger a_k \rangle$ in each cavity. Then, we discuss the effect of the number of photons in the Fock states appearing in the quantum state (\ref{psidt}).  To obtain the average number of photons $\langle  a_k^\dagger a_k \rangle$,  we use equation (\ref{Invpop}) by replacing $\sigma_z$ with the number operator $a_k^\dagger a_k$. Doing this we have  general expressions for $\langle a_k^\dagger a_k \rangle$ for any type of cavities, for example identical or non-identical cavities, in any regime, for example in the resonant, non-resonant or dispersive regime, and in any state of the control qubit, that is at any angle $\theta$.  However, for the sake of simplicity, we focus only in the  resonant case, i.e., $\Delta_0=\Delta_1=0$. For a non-identical cavities, a  straightforward calculation shows that the coefficients in (\ref{Invpop}) for the number operator $  a_k^\dagger a_k$ are explicitly given by
\begin{eqnarray}
 \langle  T_{jj}^\dagger  a_j^\dagger a_j T_{jj}\rangle &=& \sin ^2\left(g_jt \sqrt{n_j+1} \right)+n_j , \cr
 \langle T_{ii}^\dagger  a_j^\dagger a_j T_{ii} \rangle &=&n_j   , \cr
{\rm Re} [ \langle T_{ij}^\dagger  a_i^\dagger a_i  T_{ii}\rangle] &=& \cos \left(g_jt_m \sqrt{n_j+1} \right) \left[ \sin \left( g_i t \sqrt{n_i+1} \right) \sin \left( g_i \left(t-t_m\right) \sqrt{n_i+1} \right)+n_i \cos \left(g_i t_m \sqrt{n_i+1} \right)\right], \cr
{\rm Re} [ \langle T_{ij}^\dagger   a_j^\dagger a_j T_{ii}\rangle] &=& n_j \cos \left(g_i t_m \sqrt{n_i+1} \right) \cos \left(g_j t_m \sqrt{n_j+1} \right) ,\cr
\langle T_{ij}^\dagger  a_i^\dagger a_i T_{ij}\rangle&=& \sin ^2\left[g_i \left(t-t_m\right) \sqrt{n_i+1} \right]+n_i, \cr
\langle T_{ij}^\dagger   a_j^\dagger a_j T_{ij}\rangle&=& \sin ^2\left(g_j t_m \sqrt{n_j+1} \right)+n_j,
\end{eqnarray}
for $i\neq j$. Substituting these coefficients in  (\ref{Invpop}) one obtains the average number of photons $\langle a_j^\dagger a_j \rangle $ as a function of $\theta, n_j$, and $g_j$ at different times $t$ and $t_m$.  Explicitly, for the $j$-th cavity one finds
\begin{equation}\label{n0nm}
\footnotesize
\begin{array}{ll}
\langle a_j^\dagger a_j \rangle =\\%
 \hspace*{1.2cm}\displaystyle n_j+\frac{4 \left(\eta_{j} \sin ^2\left(g_j \sqrt{n_j+1} \left(t-t_m\right)\right)+\cos ^6(\theta+3\pi j/2 ) \sin ^2\left(g_j \sqrt{n_j+1} t\right)+ \eta_{j\oplus1}I(g_j,g_i,n_j,n_i,t,t_m)\right)}{2 \sin ^2(2 \theta ) \cos \left(g_j \sqrt{n_j+1} t_m\right) \cos \left(g_i \sqrt{n_i+1} t_m\right)+\cos (4 \theta )+3}, \cr
\end{array} 
\end{equation}
where $\eta_0=\sin ^4\theta \cos ^2\theta $, $\eta_1=\sin ^2\theta \cos ^4\theta $ and
\begin{equation}
I(g_j,g_i,n_j,n_i,t,t_m)=\sin ^2\left(g_j \sqrt{n_j+1} t_m\right)+2 \sin \left(g_j \sqrt{n_j+1} t\right) \sin \left(g_j \sqrt{n_j+1} \left(t-t_m\right)\right) \cos \left(g_i \sqrt{n_i+1} t_m\right).
\end{equation}
In the case of no superposition, the average number of photons reads 
\begin{equation*}
 \langle  a_j^\dagger a_j \rangle = \langle  T_{jj}^\dagger  a_j^\dagger a_j T_{jj}\rangle=-\frac{1}{2} \cos \left(2 t g_j \sqrt{ n_j+1}\right)+n_j+\frac{1}{2},
\end{equation*}
where $j=0$ ($=1$) for $\theta = 0$ ($=\pi/2$). For identical cavities, $g_0=g_1=g$ and $n_0=n_1=n$, at maximum indefiniteness, i.e., $\theta=\pi/4$,  each cavity has the same average number of photons and it is given by
\begin{equation}\label{aan}
\langle a_j^\dagger a_j \rangle = n-\frac{\cos \left[2 g \sqrt{n+1} \left(t-t_m\right)\right] +\cos \left(2 g \sqrt{n+1} t\right)-2}{2 \left[\cos \left(2 g \sqrt{n+1} t_m\right)+3\right]},
\end{equation}
From this equation we see that $n \leq \langle  a_j^\dagger a_j \rangle <(n+1)$.  Figure \ref{anp} shows the average number of photons for identical cavities, i.e., $g_0=g_1=g$ and $n_0=n_1=n$  for several values of $\theta$ and $n$ using equation \ref{n0nm}. For the case of zero photons, $n=0$, notice that $ \langle  a_j^\dagger a_j \rangle= 1/2$ for $\theta=\pi/4$, while the average photon number oscillates for $\theta \neq \pi/4$. For the case $n_0=n_1=10$, the average photon number is always oscillating with the minimum amplitude at the maximum indefiniteness. Notice that  oscillations of the average number of photons are equal in both cavities when the control parameter $\theta$ correspond to one cavity, i.e.   $\theta = 0$ for cavity 0 or $\theta=\pi/2$ for cavity 1 respectively.  For different weights in the superposition of both cavities, the average number  of each cavity is different even if both cavities have the same number of photons of the field.  We see also that the amplitude of the oscillations in the average number of photons is always lower in the case of  superimposed cavities than cavities  with no superposition. 
 Figure \ref{noma} shows the average number of photons $\langle  a_j^\dagger a_j\rangle$   from equation (\ref{aan}) as function of the measurement time $t_m$ for a given $t$ and $n_0=n_1=1$.  
\subsubsection*{The quantum state of the system}
\begin{figure}[htbp]
\centering
\begin{tabular}{cc}
\includegraphics[scale=0.45]{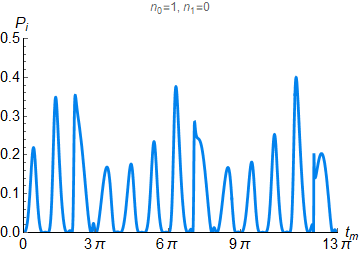}&  \includegraphics[scale=0.45]{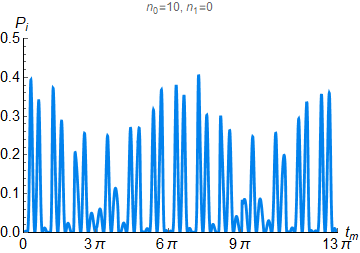}\\(a) &(b)\\
\includegraphics[scale=0.45]{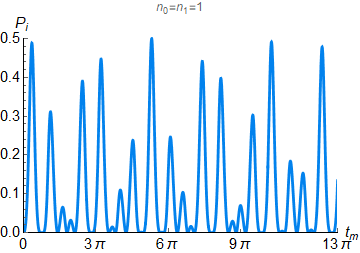}&  \includegraphics[scale=0.45]{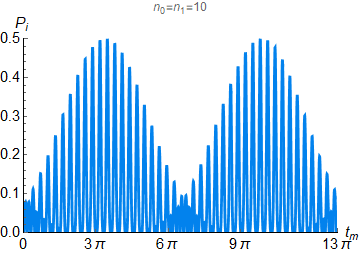}\\(c) &(d)\\
\end{tabular}
\caption{Total  probability  $ \mathcal{P}_i $  to interchange one photon between two cavities as a function of $t_m$ for two different cases.  Case 1: non-identical cavities. Figure (a) with  $n_0=1$ and $n_1=0$  and Figure (b) with $n_0=10$ and $n_1=0$. Case 2: identical cavities. Figure (c) with $n_0=1=n_1=1$ and Figure (d) with $n_0=1=n_1=10$. All plots were done for a given time $t=64/5 \pi$ and coupling parameter $g_1=g_2=0.2$.}
\label{pro}
\end{figure}

\begin{figure}[htbp]
\centering
\begin{tabular}{cc}
\includegraphics[scale=0.5]{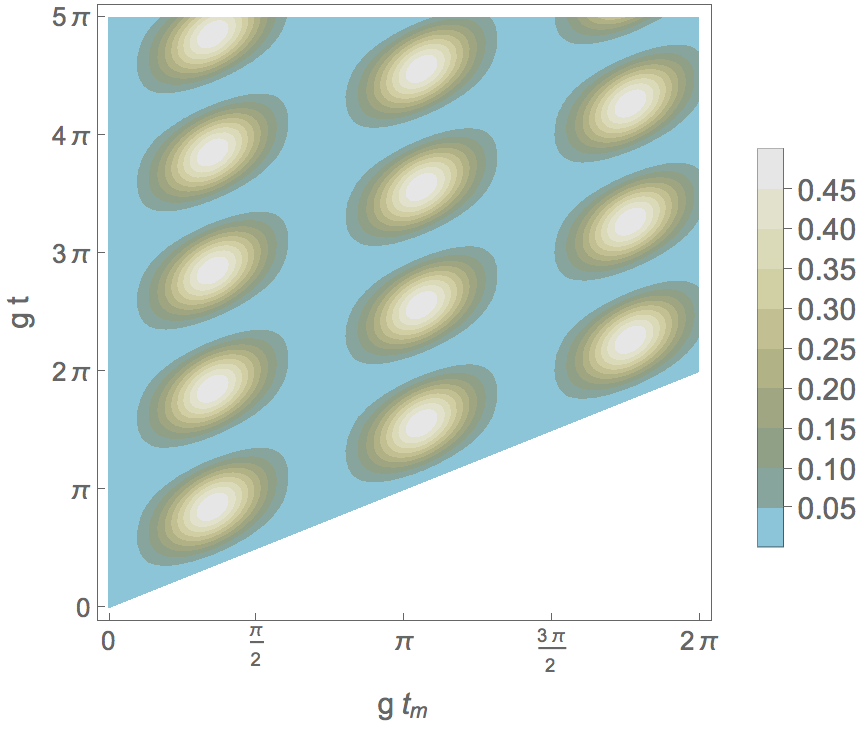}\\
\end{tabular}
\caption{Total probability $\mathcal{P}_i $ , from equation (\ref{ptn}), to interchange one photon between two identical cavities at maximum indefiniteness as function of the measurement $t_m$ and $t$ for $n_0=n_1=1$ and $g_0=g_1=0.2.$}
\label{probattm}
\end{figure}

\vspace{0.5ex}
To analyze the states in  (\ref{psidt}), we calculate the action of operators $T_{ij}$  on the state $\vert e,n_0,n_1\rangle$ without assuming any type of regime or cavities.  The most general quantum state (\ref{psidt}) of the system at time $t$ is
\begin{eqnarray}\label{estado}
 \vert \psi(t)\rangle&=& \frac{\vert 0\rangle}{{\cal N}_0}\otimes(\xi_1  \vert e,n_0,n_1\rangle+\xi_2  \vert g,n_0+1,n_1\rangle +\xi_3  \vert g,n_0,n_1+1\rangle+ \xi_4  \vert e,n_0-1,n_1+1\rangle) \cr
&& +\frac{\vert 1\rangle}{{\cal N}_0} \otimes(\xi_5  \vert e,n_0,n_1\rangle + \xi_6  \vert g,n_0,n_1+1\rangle+ \xi_7  \vert g,n_0+1,n_1\rangle+\xi_8  \vert e,n_0+1,n_1-1\rangle) e^{i\varphi}.
\end{eqnarray}
where $\xi_k := \xi_k(g_0,g_1,n_0,n_1,\Delta_0,\Delta_1,\omega_0,\omega_1,t,t_m)$, for $k=1,2,\ldots,8$. Explicitly,
\begin{equation}
\begin{array}{ll}
 \xi_1 = \cos^3 \theta e^{\phi_0'} e^{\phi_0} f_c(n_0+1,\Delta_0,g_0,t) + \cos \theta\sin^2 \theta e^{\phi_0'}e^{\phi_1}f_c (n_0+1,\Delta_0,g_0,t-t_m)f_c (n_1+1,\Delta_1,g_1,t_m) \\
\xi_2 = -\cos \theta\sin^2 \theta   e^{\phi_0'} e^{\phi_1}  h_c (n_0+1,\Delta_0,g_0,t-t_m)f_c (n_1+1,\Delta_1,g_1,t_m)-h_c(n_0+1,\Delta_0,g_0,t) \cos^3 \theta e^{\phi_0'} e^{\phi_0}, \\
\xi_3 =-\cos \theta\sin^2 \theta e^{\phi_0'} e^{\phi_1} h_c (n_1+1,\Delta_1,g_1,t_m)f_c^* (n_0,\Delta_0,g_0,t-t_m),\\
\xi_4 =\cos \theta\sin^2 \theta e^{\phi_0'} e^{\phi_1} h_c (n_1+1,\Delta_1,g_1,t_m)h_c (n_0,\Delta_0,g_0,t-t_m) ,\\
\xi_5 = e^{i\varphi} \cos^2 \theta\sin \theta e^{\phi_1'} e^{\phi_0}  f_c (n_0+1,\Delta_0,g_0,t_m)f_c (n_1+1,\Delta_1,g_1,t-t_m) + e^{i\varphi} \sin^3 \theta e^{\phi_1'} e^{\phi_1}  f_c (n_1+1,\Delta_1,g_1 ,t), \\
\xi_6=-e^{i\varphi} \cos^2 \theta\sin \theta e^{\phi_1'} e^{\phi_0}  f_c (n_0+1,\Delta_0,g_0,t_m)h_c (n_1+1,\Delta_1,g_1,t-t_m) - e^{i\varphi} \sin^3 \theta e^{\phi_1'} e^{\phi_1}  h_c (n_1+1,\Delta_1,g_1 ,t),\\
\xi_7= -e^{i\varphi} \cos^2 \theta\sin \theta e^{\phi_1'} e^{\phi_0}  h_c (n_0+1,\Delta_0,g_0,t_m)f_c^* (n_1,\Delta_1,g_1,t-t_m)  ,\\
\xi_8=e^{i\varphi} \cos^2 \theta\sin \theta e^{\phi_1'} e^{\phi_0}  h_c (n_0+1,\Delta_0,g_0,t_m)h_c (n_1,\Delta_1,g_1,t-t_m) ,
\end{array}
\end{equation}
where 
\begin{equation*}
f_c (n_j,\Delta_j,g_j,t)=\cos (t\Omega_{j}(n_j)) -\frac{1}{2} i \Delta _j \sin (\Omega_{j}(n_j)t)/\Omega_{j}(n_j),   \quad  h_c (n_j,\Delta_j,g_j,t)=-i g_j \sqrt{n_j} \sin (t\Omega_{j}(n_j))/\Omega_{j}(n_j),
\end{equation*}  
and the operators $e^{\phi_j} =e^{-i\omega_{j}(n_{j}+ 1/2)t_m} e^{-i\omega_{j\oplus1}a_{j\oplus1}^{\dagger}a_{j\oplus1}t_m}$ and $e^{\phi_j'}=e^{-i\omega_{j}(n_{j}+ 1/2)(t-t_m)} e^{-i\omega_{j\oplus1}a_{j\oplus1}^{\dagger}a_{j\oplus1}(t-t_m)}$. Besides, the coefficients $\xi_i's$ satisfy $ \frac{1}{{\cal N}_0^2} \sum_{j=1}^{8} \vert  \xi_j \vert^2=1$. Notice that, for the resonant case, i.e., $\Delta_0=\Delta_1=0$, the phases $e^{\phi_j}$ will be cancelled when calculating the probabilities  $\vert  \xi_j \vert^2 $.  From (\ref{estado})  we observe that states $ \vert e,n_0-1,n_1+1\rangle$  and $ \vert e,n_0+1,n_1-1\rangle $ describe the atom as an intermediary to pass a photon from one cavity to the other in such a way that the atom remains in the excited state. Thus, the probability of finding the atom in the excited state, the cavity 0 with $n_0-1$ photons and  the cavity 1 with $n_1+1$ photons is  $\vert \xi_4 \vert^2/{\cal N}_0^2$, while the probability of finding the atom in the excited state, cavity 0 with $ n_0+1$ photons and cavity 1 with $n_1-1$ photons is $\vert \xi_8 \vert^2/{\cal N}_0^2$. For the case of non-identical cavities, the total  probability  $\mathcal{P}_i =\frac{\vert \xi_4 \vert^2}{{\cal N}_0^2}+\frac{\vert \xi_8 \vert^2}{{\cal N}_0^2}$  to interchange one photon between, in the case of resonance and at the maximum indefiniteness, i.e. $\theta=\pi/4$,  is found to be
\begin{equation}
\mathcal{P}_i  =\frac{\sin ^2\left[g_1 \sqrt{n_1} \left(t-t_m\right)\right] \sin ^2\left(g_0 \sqrt{n_0+1} t_m\right)+\sin ^2\left[ g_0 \sqrt{n_0} \left(t-t_m\right)\right] \sin ^2\left(g_1 \sqrt{n_1+1} t_m\right)}{4 \cos \left(g_0 \sqrt{n_0+1} t_m\right) \cos \left(g_1 \sqrt{n_1+1} t_m\right)+4}
\end{equation}
For the case of identical cavities, i.e., $g_0=g_1=g$ with  $n_0=n_1=n$, the probability $\mathcal{P}_i $ reduces to
\begin{equation}\label{ptn}
\mathcal{P}_i =\frac{\sin ^2\left(g \sqrt{n} \left(t-t_m\right)\right) \sin ^2\left(g \sqrt{n+1} t_m\right)}{ \cos \left(2 g \sqrt{n+1} t_m\right)+3} ,
\end{equation}
which achieves a maximum value $\mathcal{P}_i =0.5$ when $t_m=\frac{\pi  (2 l-1)}{2 g \sqrt{n+1}}$ and $t=\frac{\pi}{2 g}  \left(\frac{1}{\sqrt{n}}+\frac{1}{\sqrt{n+1}}\right) (2 l-1)$ for any integer $l$, any number of photons $n$, and any value of $g$.  Figure \ref{pro} illustrates $\mathcal{P}_i$  for different types of cavities. For non-identical cavities, Figure 6(a) shows $\mathcal{P}_i$ with $n_0=1$ and $n_1=0$ as a function of $t_m$, while Figure 6(b) shows $\mathcal{P}_i$ with $n_0=10$ and $n_1=0$. For identical cavities, Figure 6(c) shows $\mathcal{P}_i$ with $n_0=n_1=1$ as a function of $t_m$, while Figure 6(d)  shows $\mathcal{P}_i$ with $n_0=n_1=10$.  We can see that  probability to interchange the photon presents an envelope whose maximum value is 0.5 for the case of identical number of photons, $n_0=n_1=n$, while $\mathcal{P}_i <0.5$ when the cavities initially contain a different number of photons.  Figure \ref{probattm} shows  $\mathcal{P}_i $    from equation (\ref{ptn}) as function of the measurement $t_m$ and a given time $t$ for $n_0=n_1=1$. 
%%%
%%%
%%%

\section{The two cavity system in the dispersive regime}\label{dispersiveR}

The objective of this section is to determine what type of entangled states of the two cavity fields can be created and if it is possible to prepare each cavity field in a Schr\"{o}dinger cat state when the atom interacts dispersively with both cavity fields. For example, in both the JC \cite{Haroche} and Rabi models \cite{PRA,PRAa} it is well known that Schr\"{o}dinger cat states can be created in the cavity field when it interacts dispersively with the atom. We continue to use the notation introduced in the previous two sections and the quantities have exactly the same meaning.

The setup has been described in Fig. ~ \ref{Figure0}(b). At time $t = 0$ the state of the complete system is prepared and the atom is shot towards the cavities. It moves with constant velocity, enters the cavities at a time $t = T_{0} \geq 0$, and then exits them at a time $t= T_{0} + t_{m}$ with $t_{m} = (2m-1)\pi\vert \Delta \vert / (2g^{2})$ for some positive integer $m$. Afterwards, at a time $t \geq T_{0} + t_{m}$ the state of the control qubit and the state of the atom are measured in succession.

The Hamiltonian of the complete system is 
\begin{eqnarray}
\label{39}
H(t) &=&
\left\{
\begin{array}{cc}
H_{\mbox{\tiny free}} & \mbox{if $0\leq t < T_{0}$}, \cr
H_{I} & \mbox{if $T_{0} \leq t \leq T_{0} + t_{m}$}, \cr
H_{\mbox{\tiny free}} & \mbox{if $T_{0} + t_{m} < t$}.
\end{array}
\right.
\end{eqnarray}
where $H_{\mbox{\tiny free}}$ and $H_I$ are given by equations (\ref{hfree}) and (\ref{hinteraction}), respectively. 

In all that follows we assume that the cavities are identical, that is, $\omega_{0} = \omega_{1} = \omega$ and $g_{0} = g_{1} = g >0$. In addition, we shall be working in the \textit{linear dispersive regime} \cite{Blais,Zueco}, that is, we assume that 
\begin{eqnarray}
\label{regimen}
\lambda \equiv \frac{g}{\vert \Delta \vert} &\ll& 1, \cr 
\lambda^{2}(n_{\mathrm{max}}+1) &\leq& 10^{-2},
\end{eqnarray}
where the dynamics of cavity field $k$ are approximately restricted to the subspace spanned by $\{ \vert n \rangle_{k}: \ n=0,1,2,\dots, n_{\mathrm{max}} \}$ for some positive integer $n_{\mathrm{max}}$ ($k=0,1$). Since the atom can only add one photon to the cavity fields, the value of $n_{\mathrm{max}}$ can be estimated by $n_{\max} = \mathrm{max}_{k=0,1}[\langle a_{k}^{\dagger} a_{k} \rangle(0) + 10\Delta(a_{k}^{\dagger}a_{k})(0)]$ where $\langle a_{k}^{\dagger} a_{k} \rangle$ is the expected value of the number of photons in cavity field $k$ at time $t=0$ and $\Delta(a_{k}^{\dagger}a_{k})(0)$ is its standard deviation. For example, $n_{\mathrm{max}} = 100$ requires $\lambda \lesssim 0.01$, while $n_{\mathrm{max}} = 28$ needs $\lambda \leq 0.019$. Under these conditions one can approximate $H_{\mbox{\tiny JC}}^{(k)}$ by the \textit{linear dispersive JC Hamiltonian} \cite{Blais,Zueco}
\begin{eqnarray}
\label{42h}
H_{\mbox{\tiny JCD}}^{(k)} &=& \frac{\hbar}{2}\left( \omega_{a} + \Delta \lambda^{2} \right) \sigma_{z} + \hbar\left( \omega + \Delta \lambda^{2} \sigma_{z} \right) a_{k}^{\dagger} a_{k} + \hbar \frac{\Delta \lambda^{2}}{2}  \ , \quad (k=0,1).
\end{eqnarray}

Assume that the initial state of the complete system is a separable state of the form
\begin{eqnarray}
\label{40}
\vert \psi (0) \rangle &=& \vert + \rangle_{c} \otimes \frac{1}{\sqrt{2}}\Big( \vert g \rangle + e^{i\chi}\vert e \rangle \Big)  \otimes \vert \alpha \rangle_{0} \otimes \vert \alpha \rangle_{1}, 
\end{eqnarray}
where the state of the control qubit is given in Eq.~(\ref{Control}), $\chi$ is a real number, and $\vert \alpha \rangle_{k}$ with $k=0,1$ denotes a coherent state of cavity field $k$.
One requires that the state of the control qubit is $\vert + \rangle_{c}$ so that the atom passes through both cavities. The objective of this is to have a situation similar to Young's double slit experiment. 

The state of the system at a time $t \geq T_{0} + t_{m}$ is given by
\begin{eqnarray}
\vert \psi (t) \rangle
&=& e^{-\frac{i}{\hbar}H_{\mbox{\tiny free}}(t-T_{0}-t_{m})} e^{-\frac{i}{\hbar}H_{I}t_{m}} e^{-\frac{i}{\hbar}H_{\mbox{\tiny free}}T_{0}} \vert \psi (0) \rangle \cr
&=& \frac{e^{i\omega_{a} t/2}}{\sqrt{2}}\Bigg[
\vert 0 \rangle_{c} \otimes \vert \psi_{0}(t) \rangle \otimes \vert \alpha e^{-i\omega t} \rangle_{1} 
 \ + \vert 1 \rangle_{c} \otimes \vert \alpha e^{-i\omega t} \rangle_{0} \otimes \vert \psi_{1}(t) \rangle   
\Bigg] \ ,
\end{eqnarray}
with
\begin{eqnarray}
\label{45c}
\vert \psi_{k}(t) \rangle &=& \frac{1}{\sqrt{2}} \Big[ \vert g \rangle \otimes \vert - \alpha_{m}(t) \rangle_{k} + i(-1)^{m} e^{i(\chi -\omega_{a} t)} \vert e \rangle \otimes \vert \alpha_{m}(t) \rangle_{k} \Big] , \cr
\alpha_{m}(t) &=& i(-1)^{m}\alpha e^{-i\omega t} , \quad (k=0,1). 
\end{eqnarray}
This result was obtained by using that 
\begin{eqnarray}
\label{accion}
e^{-\frac{i}{\hbar}H_{I}t_{m}} &=& e^{-\frac{i}{\hbar}H_{\mbox{\tiny JCD}}^{(0)}t_{m}} e^{- i \omega a_{1}^{\dagger}a_{1} t_{m}} \vert 0 \rangle_{cc}\langle 0 \vert+ e^{-\frac{i}{\hbar}H_{\mbox{\tiny JCD}}^{(1)}t_{m}} e^{-i \omega a_{0}^{\dagger}a_{0} t_{m}} \vert 1 \rangle_{cc}\langle 1 \vert \ .
\end{eqnarray}

Now fix a time $t \geq T_{0} + t_{m}$. This corresponds to any time after the atom has exited the cavities. First, measure at time $t$ the state of the control qubit to see if it is in the state $\vert + \rangle_{c}$ or $\vert - \rangle_{c}$. Immediately afterwards, measure the state of the atom. In order to express the results succinctly it is convenient to define the following normalized states for each $k=0,1$:
\begin{eqnarray}
\label{definiciones}
\vert \mathrm{cat} \rangle_{k} &=& \frac{1}{\mathcal{N}} \Big[ \ \vert - \alpha_{m}(t) \rangle_{k} + i (-1)^{m}e^{i(\chi - \omega_{a} t)} \vert \alpha_{m}(t) \rangle_{k} \ \Big] \ , \cr
\vert \uparrow_{\pm} \rangle_{k} &=& \vert \pm \alpha_{m}(t) \rangle_{k} \ , \cr
\vert \downarrow \rangle_{k} &=& \vert \alpha e^{-i\omega t} \rangle_{k} \ ,  \cr
\vert \mbox{Bell}_{\pm} \rangle &=& \frac{1}{\mathcal{N}_{\mbox{\tiny Bell}}}\Big[ \vert \uparrow_{\pm} \rangle_{0} \otimes \vert \downarrow \rangle_{1} + \vert \downarrow \rangle_{0} \otimes \vert \uparrow_{\pm} \rangle_{1} \Big] \ , \cr
\vert \mbox{bell}_{\pm} \rangle &=& \frac{1}{\mathcal{N}_{\mbox{\tiny bell}}}\Big[ \vert \uparrow_{\pm} \rangle_{0} \otimes \vert \downarrow \rangle_{1} - \vert \downarrow \rangle_{0} \otimes \vert \uparrow_{\pm} \rangle_{1} \Big] \ .
\end{eqnarray}
The normalization constants are given by

\begin{eqnarray}
\label{definicionesb}
\mathcal{N} &=& \sqrt{2} \sqrt{1 - (-1)^{m}e^{-2\vert \alpha \vert^{2} } \mbox{sin}(\chi - \omega_{a} t)} \ ,  \cr
\mathcal{N}_{\mbox{\tiny{Bell}}} &=& \sqrt{2}\sqrt{1 + e^{-2\vert \alpha \vert^{2}}} \ , \cr
\mathcal{N}_{\mbox{\tiny bell}} &=& \sqrt{2}\sqrt{1 - e^{-2\vert \alpha \vert^2}} \ .
\end{eqnarray}
Observe that $\vert \mathrm{cat} \rangle_{k}$ is a Schr\"{o}dinger cat state for cavity field $k$ and that the overlap of the coherent states composing it is $\vert_{k} \langle -\alpha_{m}(t) \vert \alpha_{m}(t) \rangle_{k} \vert = e^{-2\vert \alpha \vert^{2}}$. Hence, the cat state is easily distinguishable if $[e^{-2\vert \alpha \vert^{2}} \leq 10^{-2}\  \Leftrightarrow \vert \alpha \vert^{2} \geq \mbox{ln}(10) = 2.3]$. Therefore, the initial states $\vert \alpha \rangle_{k}$ of the cavity fields require an expected photon number $_{k}\langle \alpha \vert a_{k}^{\dagger}a_{k} \vert \alpha \rangle_{k} = \vert \alpha \vert^{2} \geq 2.3$ to have well defined cat states.

The notation $\vert \uparrow_{\pm} \rangle_{k}$ and $\vert \downarrow \rangle_{k}$ was introduced to suggest a similarity with qubit Bell states \cite{Barnett}. Here field coherent states play the role of the excited and ground states of a qubit. The overlap between $\vert \uparrow_{\pm} \rangle_{k}$ and $\vert \downarrow \rangle_{k}$ is $\vert_{k} \langle \downarrow \vert \uparrow_{\pm} \rangle_{k} \vert =  e^{-\vert \alpha \vert^{2}}$.
Hence,  $\vert \uparrow_{\pm} \rangle_{k}$ and $\vert \downarrow \rangle_{k}$ are approximately orthogonal if $[e^{-\vert \alpha \vert^{2}} \leq 10^{-2} \Leftrightarrow \vert \alpha \vert^{2} \geq 2\mbox{ln}(10) = 4.6]$. Under this condition the states $\vert \mbox{Bell}_{\pm} \rangle$ and $\vert \mbox{bell}_{\pm} \rangle$ have a form similar to the Bell states of a qubit. In the following, whenever discussing these \textit{cavity fields' Bell states} we shall assume that the expected number of photons in the initial states of the cavity fields is $_{k}\langle \alpha \vert a_{k}^{\dagger}a_{k} \vert \alpha \rangle_{k} = \vert \alpha \vert^{2} \geq 4.6$.

Finally, observe that $\vert_{k} \langle \downarrow \vert \mathrm{cat} \rangle_{k} \vert \leq 2e^{-\vert \alpha \vert^{2}}/\mathcal{N}$. Hence, $\vert \downarrow \rangle_{k}$ and $\vert \mathrm{cat} \rangle_{k}$ are approximately orthogonal if $\vert \alpha \vert^{2} \geq 5$ because the overlap is $\vert_{k} \langle \downarrow \vert \mathrm{cat} \rangle_{k} \vert < 10^{-2}$. 

From what has been presented in the paragraphs above, it is sufficient to consider initial states $\vert \alpha \rangle_{k}$ of the cavity fields such that the expected number of photons is $_{k}\langle \alpha \vert a_{k}^{\dagger}a \vert \alpha \rangle_{k} = \vert \alpha \vert^{2} \leq 5$. Hence, one can take, for example, $n_{\max} = 28$ because $\mathrm{max}_{k=0,1}[_{k}\langle \alpha \vert a_{k}^{\dagger} a_{k} \vert \alpha \rangle_{k} + 10\Delta(a_{k}^{\dagger}a_{k})] = \vert \alpha \vert^{2} + 10\vert \alpha \vert = 27.4$.
%%%%%%%%%

\subsection{Control qubit in the state $\vert + \rangle_{c}$}

In this and only this section assume that the control qubit is found in the state $\vert + \rangle_{c}$. Then, the state of the complete system immediately after the measurement is 
\begin{eqnarray}
\label{50}
\vert \psi_{\mbox{\tiny M}} \rangle &=& \frac{1}{\sqrt{2}}\vert + \rangle_{c} \otimes \Bigg[ \vert g \rangle \otimes \vert \mbox{Bell}_{-} \rangle + i(-1)^{m}e^{i(\chi - \omega_{a} t)}\vert e \rangle \otimes \vert \mbox{Bell}_{+} \rangle \Bigg] .
\end{eqnarray}
If immediately after the measurement of the control qubit one measures the state of the atom to see if it is in the excited $\vert e \rangle$ or ground $\vert g \rangle$ state, then the cavity fields will be prepared in one of the highly entangled $\vert \mbox{Bell}_{\pm} \rangle$ states.  

Now consider the case where, immediately after the measurement of the control qubit, one measures the state of the atom to see if it is in the  $\vert + \rangle_{x} = (1/\sqrt{2})(\vert e \rangle + \vert g \rangle )$ or $\vert - \rangle_{x} = (1/\sqrt{2})(\vert e \rangle + \vert g \rangle )$ state. 

Assume that the atom is found in the $\vert + \rangle_{x}$ state. If the atom is found in the $\vert - \rangle_{x}$ state, then one only needs to replace $\chi$ by $( \chi+\pi )$ and $\vert + \rangle_{x}$ by $\vert - \rangle_{x}$ in the results below. The state of the complete system immediately after the measurement of the state of the atom is
\begin{eqnarray}
\label{57o}
\vert \psi_{\mbox{\tiny MM}} \rangle &=& \frac{1}{\mathcal{N}_{\mbox{\tiny MM}}} \vert + \rangle_{c} \otimes \vert + \rangle_{x} \otimes \Bigg[ \vert \mathrm{cat}  \rangle_{0} \otimes \vert \downarrow \rangle_{1} + \vert \downarrow \rangle_{0}\otimes \vert \mathrm{cat} \rangle_{1} \Bigg] \ ,
\end{eqnarray}
with $\mathcal{N}_{\mbox{\tiny MM}}$ a normalization constant. Observe that the cavity fields are in a highly entangled state that also resembles a qubit Bell state if $\vert \alpha \vert^{2} \geq 5$ because $\vert_{k} \langle \downarrow \vert \mathrm{cat} \rangle_{k} \vert < 10^{-2}$.

\begin{figure}[h]
\centering
\includegraphics[scale=0.75]{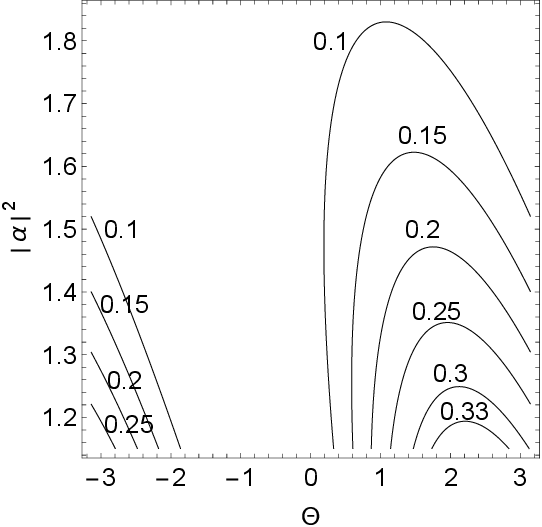}
\caption{\label{Figure5} The figure illustrates a contour plot of the probability in (\ref{59}) to find both cavity fields in a Schr\"{o}dinger cat state as a function of $\Theta$ and the expected number of photons $\vert \alpha \vert^{2}$ in the coherent state $\vert \alpha \rangle_{k}$ $(k=0,1)$. }
\end{figure}

Given that Schr\"{o}dinger cat states appear in the superposition between brackets on the righthand side of (\ref{57o}), there can be a nonnegligible probability to find each cavity field in a cat state. The probability to find the cavity fields in the state $\vert \mathrm{cat} \rangle_{0} \otimes \vert \mathrm{cat} \rangle_{1}$ immediately after the measurement of the state of the atom is 
\begin{eqnarray}
\label{59}
\mathcal{P} &=& 2\frac{1 - \mbox{sin}(\Theta + 2\vert \alpha \vert^{2})}{e^{2\vert \alpha \vert^{2}} + 1 -\mbox{sin}\Theta- \mbox{sin}(\Theta+ 2\vert \alpha \vert^{2}) } \ ,
\end{eqnarray}
with
\begin{eqnarray}
\label{59b}
\Theta &=& (-1)^{m}(\chi - \omega_{a} t) .
\end{eqnarray}
Observe that $\Theta$ and $\vert \alpha \vert^{2}$ are parameters that can be adjusted by changing the time $t\geq T_{0} + t_{m}$ when one performs the measurements and by preparing the initial state of the cavity fields. Notice that one must optimize the probability while still preserving easily distinguishable cat states. Figure \ref{Figure5} illustrates the probability as a function of these two parameters. Observe that one can achieve a probability $\mathcal{P} \lesssim 0.35$ and, in particular, that $\mathcal{P} = 0.35$ if $\Theta = 2.25$ and $\vert \alpha \vert^{2} = 1.155$. Notice that for $\vert \alpha \vert^{2} = 1.155$ one still has reasonably distinguishable cat states $\vert \mathrm{cat} \rangle_{k}$ because the overlap between the coherent states composing the cat state is $\Big\vert \langle -\alpha_{m}(t) \vert \alpha_{m}(t) \rangle \Big\vert = 0.1$.

%%%%%%----->Section
\subsection{Control qubit in the state $\vert - \rangle_{c}$}

In this and only this section assume that the control qubit is found in the state $\vert - \rangle_{c}$. Then, the state of the complete system immediately after the measurement is 
\begin{eqnarray}
\label{50}
\vert \psi_{\mbox{\tiny M}} \rangle &=& \frac{1}{\sqrt{2}}\vert - \rangle_{c} \otimes \Bigg[ \vert g \rangle \otimes \vert \mbox{bell}_{-} \rangle  + i(-1)^{m}e^{i(\chi - \omega_{a} t)}\vert e \rangle \otimes \vert \mbox{bell}_{+} \rangle \Bigg] .
\end{eqnarray}
If immediately after the measurement of the control qubit one measures the state of the atom to see if it is in the excited $\vert e \rangle$ or ground $\vert g \rangle$ state, then the cavity fields will be in one of the highly entangled $\vert \mbox{bell}_{\pm} \rangle$ states.

Now consider the case where, immediately after the measurement of the control qubit, one measures the state of the atom to see if it is in the  $\vert + \rangle_{x}$ or $\vert - \rangle_{x}$ state. 

Assume that the atom is found in the $\vert + \rangle_{x}$ state. If the atom is found in the $\vert - \rangle_{x}$ state, then one only needs to replace $\chi$ by $( \chi+\pi )$ and $\vert + \rangle_{x}$ by $\vert - \rangle_{x}$ in the results below. The state of the complete system immediately after the measurement of the state of the atom is
\begin{eqnarray}
\label{57}
\vert \psi_{\mbox{\tiny MM}} \rangle &=& \frac{1}{\mathcal{N}_{\mbox{\tiny MM}}} \vert - \rangle_{c} \otimes \vert + \rangle_{x} \otimes \Bigg[ \vert \mathrm{cat}\rangle_{0} \otimes \vert \downarrow \rangle_{1}  - \vert \downarrow \rangle_{0} \otimes \vert \mathrm{cat} \rangle_{1} \Bigg] \ ,
\end{eqnarray}
with $\mathcal{N}_{\mbox{\tiny MM}}$ a normalization constant. It follows that the cavity fields are in a highly entangled state that has the form of a qubit Bell state if $\vert \alpha \vert^{2} \geq 5$ because $\vert_{k} \langle \downarrow \vert \mathrm{cat} \rangle_{k} \vert < 10^{-2}$. In this case the probability to find the cavity fields in the state $\vert \mathrm{cat} \rangle_{0} \otimes \vert \mathrm{cat} \rangle_{1}$ immediately after the measurement of the state of the atom is zero due to the minus sign in the linear combination of states inside the brackets in Eq.~(\ref{57}).

\section{Conclusions}

In this article we studied the effects of causal indefiniteness in a cavity quantum electrodynamics setup where an atom passes at the same time through two cavities by using a control qubit. Moreover, measurements are performed on the control qubit and the atom. Two scenarios were considered. In the first one, the atom interacts resonantly with both cavity fields which are initially prepared in Fock states. The dynamics of the system were considered while the atom is inside the cavities and it was found that the atom can function as a \textit{shuttle} that can send a photon from one cavity to the other without changing its state. Moreover, it was determined that the Rabi oscillations can be modified to have a smaller amplitude or a \textit{beats structure} similar to that of two resonantly coupled harmonic oscillators. In the second scenario the atom interacts dispersively with both cavity fields which are initially prepared in a coherent state. The generation of entanglement between the two cavity fields was considered once the atom exits both cavities by performing successive projective measurements on both the control qubit and the atom. It was found that the cavity fields can be left in a highly entangled state that can have the form of qubit Bell states were the excited and ground states of the qubit are replaced by approximately orthogonal field coherent states. Moreover, it was also determined that there can be a nonnegligible probability $\leq 0.35$ to find both cavity fields in a Schr\"{o}dinger cat state.

\section{Acknowledgements}
Marco Enr\'iquez and Luis Octavio Castan\~os are grateful for the support to publish this article provided by the School of Engineering and Science, Tecnologico de Monterrey. The support of CONACyT is acknowledged as well.
Lorenzo M. Procopio acknowledges the support of Israel Science Foundation. 

%\section{Author contributions statement}
%
%All authors reviewed and contributed  equally to the manuscript. 
%
%\section{Ethics declarations}
%Competing interests
%The authors declare no competing financial interests.

\end{document}